# Integration of Security Modules in Software Development Lifecycle Phases

1st Isaac Chin Eian
School of Computer Science & Engineering, Taylor's University
Selangor, Malaysia
zackteddy39@gmail.com

2nd Lim Ka Yong
School of Computer Science & Engineering, Taylor's University
Selangor, Malaysia
limkayong001117@gmail.com

3rd Majesty Yeap Xiao Li
School of Computer Science & Engineering, Taylor's University
Selangor, Malaysia
majesty2910@gmail.com

4th Noor Affan Bin Noor Hasmaddi
School of Computer Science & Engineering, Taylor's University
Selangor, Malaysia
stonexfrost@gmail.com

4th Fatima-tuz-Zahra
School of Computer Science & Engineering, Taylor's University
Selangor, Malaysia
fatemah.tuz.zahra@gmail.com

*Abstract*—Information protection is becoming a focal point for designing, creating and implementing software applications within highly integrated technology environments. The use of a safe coding technique in the software development process is required by many industrial IT security standards and policies. Despite current cyber protection measures and best practices, vulnerabilities still remain strong and become a huge threat to every developed software. It is crucial to understand the position of secure software development for security management, which is affected by causes such as human security-related factors. Although developers are often held accountable for security vulnerabilities, in reality, many problems often grow from a lack of organizational support during development tasks to handle security. While abstract safe coding guidelines are generally recognized, there are limited low-level secure coding guidelines for various programming languages. A good technique is required to standardize these guidelines for software developers. The goal of this paper is to address this gap by providing software designers and developers with direction by identifying a set of secure software development guidelines. Additionally, an overview of criteria for selection of safe coding guidelines is performed along with investigation of appropriate awareness methods for secure coding.

## I. INTRODUCTION

For any application, it is a developer's dream to make the user experience well and pleasant in a safe environment. However, in the reality of every software development process, developers struggle to maintain a balance between developing software and ensuring its security is not compromised. The costs of any insecure software will be skyrocketing high. The critical question is that since completely resilient software is only a myth, how does one or a team of developers create a good software that is also secure? The answer to that was introduced in the 1960's, according to Elliot and Strachan and Radford (2004), in the form of Secure Development Process.

However, before we begin diving into secure development processes, we need to ascertain the need for secure development processes. We need to look at why and how vulnerabilities and threats can be formed. It is not a surprise to see that during the live testing of an application, a plethora of threats and attacks occur showing the massive gap of vulnerabilities in the software as well as not to add more friction points between the team. It would be safe to say that without a proper security integration, the developers have failed to create a safe environment for users to utilize the software. As a result, this leads to insecurities at various levels, causing vulnerabilities to security attacks like ransomware [1], phishing [2] and many others. This can be blamed on many factors but all would narrow down to one issue; the lack of proper framework, procedures and reviews for systematic measures of security. Help developers and project managers should be the key players in ensuring this is mitigated and will consistently monitor and follow through with the development process from start to finish. Added to that, they must always base their management on a set of rules and procedures established pre-project. Thus, standard secure development processes should be set.

The Secure Development Process has always been an ongoing discussion among the developer community, mainly because it is such a challenging yet crucial component to any application. In most cases, security is completely disregarded in the early phases of the SDLC (Software Development Life Cycle). Software engineers are often unaware or less aware of the security approaches that lead to safe and functional software programs. The consequences of doing this is that with a dangerous approach such as that, the software accumulates vulnerabilities with every new step in the development life cycle and at the very end, produce a software that is highly vulnerable to security threats. Dmitri Nikolaenya [3] put it very plainly in his blog post where in "such an approach, every succeeding phase inherits vulnerabilities of the previous one, and the final product cumulates multiple security breaches. As a result, your company will have to pay through the nose to close these breaches and enhance software security in the future." [3].

The SDLC concept was created on the basis of creating software but is still flawed as mentioned before in terms of security aspects. The Secure Development Process consists of steps to build a safe environment for software to be developed, which we will dwell deeper into later on. It is important to know that the Secure Development Process is not meant to replace the SDLC but to be integrated along with it. The best practices of secure software development always begins by integrating security aspects into every SDLC phase regardless of methodology. Of course, different developers will have different ways of ushering in secure software development into their SDLC but ultimately have the same or similar processes at each stage. The main golden rule is to integrate security aspects into every phase to reduce costs on fixing vulnerabilities later.

In this paper, the aim is to see the whole picture of Software Development Life Cycle and how the security perspective fits into each and every phase. In every phase, we will seek out the important security measures that should be employed to ensure the security of the software throughout the life cycle. To do this, we will investigate many research papers available to the public, read, process, compile and provide a comprehensive guide to secure software development processes and steps at every SDLC phase. We will also dwell into the factors such as influencing the implementation as well as the security training and awareness required. We will also show and explain our methodology of handling the survey. Next, we will present and put forward our unique solution that will attempt to put out a methodology for a secure development process. Finally to close, we will conclude our survey report.

## II. Literature Review

Most of the organizations or businesses nowadays are having the intention of producing, publishing and sustaining usable software applications. However, the growing issues and business risks associated with the vulnerabilities of application have caused the increase in attention to the need to incorporate security into the process of development because they can lead to severe consequences like privacy issues [4]. So, it is necessary now to implement the secure software development process in an organization. A secure SDLC is a set of best practices which are designed to increase more security to the standard of SDLC. Security is required to be applied in every phase (Fig. 1.) of the secure software development process such as planning and requirements, architecture and design, test planning, coding, testing and results as well as release and maintenance. The development team needs to emphasize more on the security of every phase of the project instead of just emphasizing the features and functionality of the software application. The vulnerabilities of the software application can be reduced and the risk of having security issues can be decreased [5].

### A. Integration of security into SDLC

Secure SDLC is essential in the development of software applications. At every phase of SDLC, the developers need to be aware of the possible security issues that might happen. This involves incorporating protections of security which was not required before. The developers need to make sure they are coding with potential bugs in mind. This is because someone can theoretically gain unauthorized access to the source code of the software application to find the vulnerabilities and thus use those vulnerabilities to write malicious code to compromise the system of the software application. Hence, having a secure SDLC ensures that the software application is safe from hackers or any malicious user. Secure SDLC in a software application depends heavily on the strengths and limitations of the software project team and hence it is difficult to enforce a secure SDLC. It requires process updating, introducing new instruments and most significantly there will be practice transformations within the departments involved. A functionally secure SDLC is typically specific to each company and can also vary across different businesses [6].

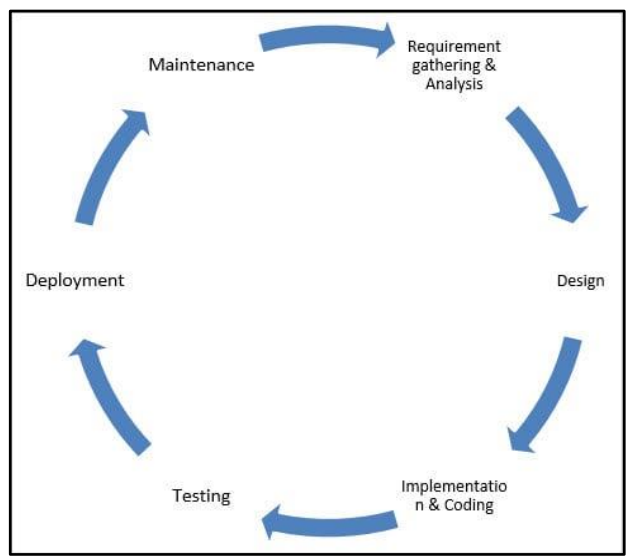

Fig. 1. Secure software development life cycle [7]

A research article conducted by TRIO company shows that there are six phases in a secure SDLC. The first phase will be planning and requirement analysis. The requirement phase is typically carried out by the more experienced senior project team members with the relevant customers' reviews and consultation and also with all the departments that are relevant such as sales and marketing. After marketing, consumer reviews, and product specifications have been collected, they will be used to prepare a project strategy and perform an analysis. Furthermore, the project manager will predict, schedule, and develop quality assurance criteria. After that, the project team will have a result from their research. When the criteria has been met by the senior project team members, they will identify and document the specifications and discuss them with the market analyst [8]. There are some practices that can be performed in this stage such as SDL discovery and preparing security requirements. It begins with identifying the security and management goals of the project. A methodology will be selected and a comprehensive plan will be created so that the project team will be concerned about the security issues in this stage. Furthermore, by preparing a list of security requirements can help the project team through identifying and fixing the non-compliant issues.

The second phase of secure SDLC is the architecture and design phase. The goal of this phase is to develop a product which can meet the criteria that is produced in the previous phase. This involves the modelling of the layout of the application and how the application is going to be used and also choosing components from the third parties which can increase the growth speed of the application. A design document will be the outcome of this phase. The practices that can be performed in this phase are threat modelling, secure design and third party-software tracking [9].

In the early stages of SDLC, some security methods are deployed. First of which is threat modeling to develop a secure system around potential threats. It focused on detecting threats so that the threats can be decreased. As the cyber security issues are more, the software project team needs to be more aware and have efficient ways to protect the software by having strong security [10]. Secondly, secure design is in the design document and changes are verified in the view of security specification. It helps to detect the features and functionalities that have the security issue risk before they are being introduced. Lastly, the third-party software tracking where the vulnerabilities of the third party items can make the system exposed to cyber security issues. Hence, it is essential to check the third party software so that the system will be safer and more secure.

The third phase of secure SDLC is implementation. In this phase, the software application is probably being created. The security activities that are focused will be secure coding requirements and code reviews. The secure coding requirement is to ensure that the developers implement a safe and secure software application. Then, the code reviews help to check if there are any security errors that have been made by the developers. There are some tools that can be used to scan the code such as Spot Bugs. These tools will provide some detailed and preventive solutions so that the quality of the code can be enhanced. It is important to do code review as the vulnerabilities of the code can be discovered and the programmers can make changes so that the attackers will have less chance to attack the system based on the vulnerabilities of the code [11].

The fourth phase of secure SDLC is the testing phase. The developers will emphasize on the analysis and research. They will figure out if the code produced and the output is based on the client specifications. If it is impossible to solve all the problems or errors that are found during this phase, the result can be used to decrease the amount of errors in the software application. In the article, it is advisable for the software project team to create a detailed test plan before they start this phase. This is because the test plan involves the testing types, resources, the ways that the software tested, the people that will be involved and also the instruction for the testing process [12]. Penetration testing can be included in this step so that the possible types of attacks can be figured out by the external experts.

The fifth phase of secure SDLC is the deployment phase. In the deployment phase, all the users are opened to the software application. There will be various platform components communicating with each other. The protection of the platform should be emphasized. This is because the application will be secure within itself. However, the software application needs to run with other platforms which might have exploitable vulnerabilities. Therefore, in order to make the platform more secure, unnecessary services should be shut off, running the devices based on the least privilege and ensuring the security protections are available such as intrusion detection systems, firewalls and others [13][14].

The last phase of secure SDLC is the maintenance phase. This phase is to ensure that if there are any problems occurred after the deployment phase and need to be solved or there are any improvements and updates need to be made, the developers will be in charge of it and make sure the software application all the configurations performed well. The security standards need to be practiced. This is to ensure the software application is safe and secure from new risks, vulnerabilities and threats. Other than that, ongoing security checks need to be performed regularly to protect the software application from any new vulnerabilities. Furthermore, any new tools also can be implemented by the developers so that the features and functionalities of the software application can be enhanced [15]. Security aspects should be integrated in all phases of the development cycle to avoid the costs which can increase exponentially. Relative defect repair costs are shown in Fig. 2.

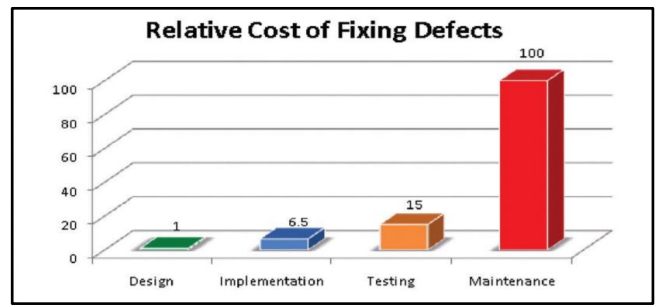

Fig. 2. Relative cost of fixing defects [16]

A discovery of vulnerability showed that the software code could involve reconstruction and implementation. It will require a lot of time and money. For a better understanding of security threats to the system, security must be implemented at the start of the development process. For a business or an organization, it means that the requirements must be enabled to guide the design and affect the software application requirements. Hence, during the requirements and analysis phase, any known security issues will be discovered. The security issues that are found can be resolved before the implementation process. By doing this, it is possible to minimize the cost within the analysing and quality checking during the phases of secure SDLC.

A research was conducted by IBM System Science Institute to identify the money that is used for the defect fixing in the secure SDLC process. The figure above has shown that the defects identified in the testing phase are 15 times more expensive than the design phase; the defects identified in the testing phase are 2 times more expensive than the implementation phase. The defects found in maintenance are most costly which is 7 times more expensive than the testing phase [16].

The growth of the internet, e-commerce and emergence of technologies like blockchain [17] and IoT has made a revolutionary change in people's life at individual as well as organizational level [18]. Today most companies and organizations run their business with operations related to online applications about its marketing activities. Web applications have become a target for malicious hackers over many years, which is why most networks must be closely monitored with and have protection measures such as firewalls or an intrusion detection system (IDS) [19]. Web application attacks are a major threat to any security software and are likely exploited from poorly constructed security systems, resulting in increased security weaknesses and defects caused by vulnerabilities that can be exploited by malicious users to violate the software security properties.

The nature of a software can have an impact on how secure development should be done as well. A security analysis can cover a wide range of approaches and the process to deal with security measures to be conducted during the software development. By following the suggested order of the SDLC process, multiple actions can be adapted to develop a secure software application. Several modifications can be made in addition to inserting traditional life cycle of security activities with security enhancements through methodologies and processes.

Other than that, software insecurity is often regarded to be caused by the negligence of security, having flaws within the software engineering process can also cause by the insufficient knowledge and understanding of the relation of the security software development. This shows how important the requirement phase is when it comes to developing any kind of software regardless of whatever nature or use it may be for. Countermeasures must be put in place to prevent such threats from occurring.

*B. Factors influencing implementation of secure software development practices*

Secure development processes incorporate security activities in each development phase with the purpose to reduce the possible number of vulnerabilities. It consists of set activities to overcome numerous security issues. This is a strategy highly used by many developers to implement a good system security and increase the effectiveness through code review, risk analysis, abuse cases, security operation and more.

The Secure Software Development (SSD) implementation can be difficult to handle as a set of predefined activities needs to be embedded into the SDLC. It can be a challenge for an organization to implement all activities and the responsibility to its team of developers. A successful implementation of using the processes is known to be done by huge organizations like Microsoft or other highly software agencies such as banks. Following common practices that are identified from the literature includes threat modelling, static analysis and security engineering requirement, it can affect the adequate development time, budget cost and security awareness. Therefore, the key categorized factors listed below describe how it will affect the SSD implementation.

*I. Developer knowledge, skills and experience:*

A successful implementation of SSD is influenced by the developer's knowledge and experience. Through experience, a developer is bound to obtain certain unique skills for the SSD [20]. Education is critical for a developer because the lack of understanding can cause a developer to unknowingly introduce vulnerability that is contained in the system. Skill can be used to implement and deploy a secure development practice that is outlined by the organization [21]. Having to acquire expensive tools by the organization, it is considered a vain effort if the developer does not have the proper skills or experience to use them. This indicates the importance for the people to be involved in the SDLC process and understand them to produce secure applications and software.

*II. Security training and awareness:*

To have effective security, an organization must conduct a training and awareness program to their developers or programmers [22]. The purpose of having the training program is to ensure the project team is armed with the necessary skills and knowledge for certain projects on the SSD implementation. The importance of awareness [23] given to every stakeholder involves ensuring support to be provided for the SSD initiative and helping developers to sensitize the potential impact of the security problem on the organization.

*III. Automated tool support:*

The SSD needs to be supported by an automated security tool. The standard usage of analysis and design will enable us to identify security strategies at design or during the requirement phase of the SDLC [24]. Despite having the developer, static analyzer, and interactive support to reduce programming errors, an automated tool is essential. It must be usable and not complex to be compatible and customized into the development environment in making sure the process will result in business value as it involves many potential stakeholders within the organization [25].

*IV. Adequate development time:*

Time is highly influential in all implantation of the SSD. This factor is highly cited in most reviews and a common reason for the development to be delayed from the actual deadline is usually due to the team ignoring practices within tight deadlines. The implementing of the SSD methodology can be time consuming and developers can be pressured into short project deadlines. An adequate time for a project must be given for the team to implement the SSD practices efficiently and meet the required deadline [26].

*V. Top management support:*

Having management support has a strong influence towards implementation of SSD practice. By obtaining top management approval and support can increase the chances of a successful implementation because it is the management responsibility to allocate the adequate resources for the project [27]. With reference [28], the article mentioned the SSD to produce a secure software, indicating that the composition of the project team along with effective collaboration and communication between members are essential for SSD implementation.

*VI. Building and retaining security team:*

Establishing and retaining a security team that will drive the implementation to a better secure development. With a well trained and skilled team, the contribution of the team expertise can enhance the system structure and make it more secure from potential threats that may occur in the future. The team must be familiar with the security updates and be consistent on the development procedures and technology [29]. This matter of expertise must be retained by the organization for the team to apply their skills on multiple different projects and given the authority to delay any project should the security requirement be not fulfilled [26], (Byers, 2007) [30].

*VII. Adequate budget cost:*

Every software initiative is influenced by the allocated costs. Depending on the implementation requirement, the adequate cost is to be considered when allocating to the project [20]. Having security tools can be very expensive. Should the tools be provided, a relevant training must be provided to the developers as well to use the tools effectively. By hiring a professional trainer for this purpose can incur cost but it is determined by the organization to overcome these constraints.

*VIII. Incentive provision for the project team:*

Providing an incentive for the team to reflect on the project development and implementation to be achieved can motivate the developers to be more productive and proactive to handle potential vulnerabilities and risk that is linked with application. Should the management desire a clean and secure code that offers perks for the application developers, can greatly achieve its goal in terms of security [31].

*IX. Security policies and reference guidelines:*

The SSD implementation can be more successful with proper security policies, standards, and guidelines. When policies are in place, the team will understand the needs for the project, providing an insight of the structure they must make and guidance to help them in their development. These references can be very helpful towards developers, especially when they are not security experts. Encouraging the team to document their experience of security for other projects [20].

*X. Clear, comprehensive and consistent security requirements:*

The SSD implementation requires determination based on the security requirements. Deriving the requirements involves the identification of its stakeholders. Security requirements must be comprehensive, clear, consistent, and unambiguous. This will allow for both stakeholders and developers to understand the security practice that must be implemented to produce the desired outcome of the project [32]. Security policies and standards can aid to determine the requirement for the project as well.

III. METHODOLOGY

Just as secure software development processes have a set of rules and procedures to follow, so must our research on this topic to ensure the content of this document is of updated material and facts. As shown above, a literature review was conducted to investigate the work that the authors had done by reviewing or providing expert opinions on secure development processes. We had utilized the existing framework of these authors as a preliminary foundation and to piece together a further study in the field. This section explains the procedures and standards taken and maintained throughout this survey to produce accurate and correct information.

The below sources were used and provided a reason to use these pieces of data, research or report. Case studies are used to explore investigations done to explain or find reason for a certain phenomenon, regarding heavily in answering the "how and why" in secure software development. Thus, we can use them as a reference point to validate research. Analysis or expert reviews by known professional project managers and software engineers can address the validity of proposed models. In most cases, these expert opinions and critiques will help build our unique solution which will be covered later on. Data models will also greatly support or deny the validity of these conclusions and methods.

Case studies are designed to investigate the stated practices and perceptions explained in the study as they relate to the appropriate requirements and importance of each SDLC phase. With this, it enables survey and research to produce complete models that will help software engineers with the necessary knowledge to adopt secure development practices and aid project managers with software security guidelines to form project management tools. For this source of information, we look at a qualitative standpoint. The overall design of study follows the qualitative paradigm as qualitative research aims to develop understanding from all angles within the complex phenomena. Primary data sources will be used only where the publication dates were not more than 5 years apart from the current year and are taken from known and prestigious publications. Secondary to that, established and prestigious authors are favoured more for their expert opinion but this is not always the case if the information is outdated or not in line with what other authors present. In most of the case studies, the main members of every team responsible were interviewed and had their input put into the case study. Direct observations and documentation analysis are also viable techniques of data collection for the study. In addition to that, multiple case studies were thoroughly read through and analyzed to gain a rich understanding of the processes being adopted and the secure frameworks.

It is important to have a standard unit of analysis in this survey. We know that software engineers and project managers play a major role in the integration of security perspectives throughout the development process. Thus, the ideas from these figures as well as the other factors that can form secure SDLC practices such as security policies, security standards, security processes, tools and the skillset of the engineers are highly looked upon. The conceptual framework (Fig. 3) shows the territory and bounds of what this survey seeks for in these research sources.

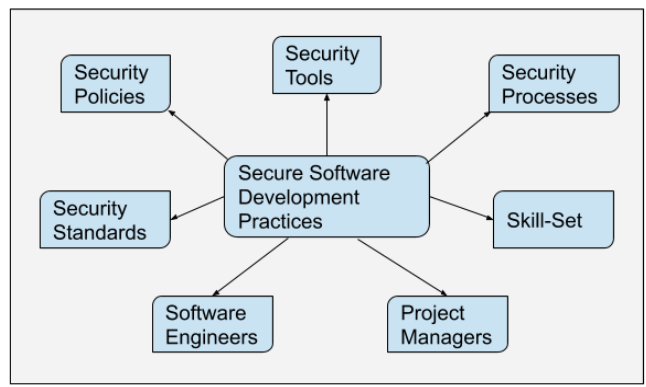

Fig. 3. Secure software development practices

Data collections are an important piece to ensure the validity and grounds of the facts explored and analyzed by the various research authors. Data on its own have no meaning unless given the right context. Therefore, it is important to understand and determine if the methodologies used to collect the data are appropriate, accurate and precise. Thorough examination and definitive decision-making will be used to acknowledge whether the data collected is correct and accurate and whether the data collected is important towards the direction of analysis. It is also important to check the timeframe of which the data was collected in which should be no more than 5 years from the current year.

Lastly, the data analysis of these authors must be acceptable or plausible enough. This is expanded upon the ideals in which the case studies were created for. Documentations, interviews, notes and structured or unstructured questionnaires can fall under this. It should be obvious to the readers that the analysis provided by them should be logical and informative, backing up their conclusions and analysis with facts from any research done and with concise and clear structuring. The analysis of prestigious and known authors are favoured as well as those that bring about the same conclusions as other authors. The analysis from these studies should not come from a single member of the team but is acceptable if it is a known author in the secure development community. Collective analysis from a team of various roles is more favourable, showing conjoined conclusions by more than one person.

IV. DISCUSSION

In a research conducted by Carleton University, security concerns, including security breaches, were faced by several participants [33]. This may be due to priority given to functionality and continuing shipping that causes the delay in security. Seven participants from the study who reported bugs in their code reported that they shipped their code with a loophole as deadlines are approaching and they decide to deal with security problems later on. This is a distinctly unsettling behavior for software development. But security problems will still be present, and it is hard to prove security completely, we can only reduce the threats and vulnerabilities [34] throughout the development process. In the following, different ways of secure development models that are suggested will be further explained in this section. One of them is USIS, which is a two-dimensional matrix with rows that lead to primary stakeholder viewpoints, while the columns list critical questions that stakeholders need to answer [35].

Other than USIS, CLASP has been an OWASP's initiative consisting of a series of modules covering the entire SDL with structured security guidelines, meaning that security concerns can be enforced from the early stages of the SDL [36]. This collection of 24 security-related activities that can be easily integrated into the SDL of the framework facilitates regular management of defense risks. This SSDL relies strongly on the coordination of the development team in areas where a precisely specified set of roles and duties can be taken care of. In comparison to most SSDLs, this is achieved where these operations are part of the SSDL's development phase. The CLASP is structured into high-level CLASP mechanism viewpoints called CLASP Views which consist of concepts view, role-based view, activity-assessment view, activity-implementation view and vulnerability view.

Apart from that, Microsoft has also developed its own methodology which is a compulsory technique used by Microsoft since 2004 to provide more secure protection and privacy applications [37]. Based on their study, design flaws were more than 50 percent of total flaws, so it is not shocking that their SDL is heavily dependent on threat modeling conducted in the early stages of development. Threat modeling is an audit technique for application security consisting of formally defining and mapping all possible vectors of attack of the submission. According to the findings provided by Microsoft, it helps decrease the number of bugs in the program code [36]. In Microsoft's SDL, it is possible to predict the probability of each hazard using the DREAD equation. The framework is decomposed into components to reveal implementation vulnerabilities and each component is evaluated according to each of the STRIDE methods where the impact of the bug is categorized using the groups with their associated opposite security properties.

Consequently, the Cigital's Software Security Touchpoints are a systematic collection of seven guiding principles proposed in 2004 that can be applied to the SDL used by the organization. Two types of program bugs are taken into account for the touchpoints: source code level and architectural level [38]. The touchpoints are categorized in order of their efficacy starting from code review, vulnerability analysis, penetration checking, risk-based compliance checks, abuse cases, security criteria and security operations. Additionally, researchers have also performed reviews and analyses [39] of various software development methodologies to for selection of best possible practices to be adopted by the developers and other related stakeholders.

V. UNIQUE SOLUTION

In this section, steps that can be taken to mitigate security vulnerabilities at each point of the software development cycle is addressed. In the initial stages of the SDLC, it is a trend for businesses offering software development and ignoring security concerns. With such a method, the following stages adopt the preceding stage's vulnerabilities, and numerous security violations are accrued by the end product that bring huge costs to the organization. Best practises in the production of software recommends incorporating security considerations into each step of SDLC, independent of project approach, from requirement analysis stage to maintenance [3].

In the requirement analysis stage, there are two main points to be taken note of. The first one is to employ a mixture of cases of use and misuse. Risk factors to the software should be foreseen by security experts and expressed in cases of abuse. Synchronously, the mitigating measures mentioned in the use cases should protect certain cases. An example of a case of misuse will be an unauthorised consumer who tries to obtain access to an application from a client and the following use case should be the system can register and evaluate any such efforts. Another key point of this stage is to carry out a risk assessment for protection and build a risk tolerance. In this case, obey recommendations from applicable authoritative outlets, such as HIPAA and SOX, when evaluating protection risks. Security experts can provide the risk profile of the software to project managers who build the project specifications at the point of requirement review. This profile includes software structures that are prone to security vulnerabilities and malware threats defined by degree of seriousness.

The design stage is followed by the requirement analysis stage, which should adopt the six security standards. The first one is least privilege feature which stated software architecture should permit regular operation with minimum user privileges. Next, it should be designed so that we can separate privilege, with less people having greater rights should be permitted to execute specific software acts. Apart from that, the software should allow full mediation with the authorization of each person using the software should be reviewed. It should also be designed with many levels of security so that every possibility of security vulnerability that would break the whole software will be removed by implementing this theory. In addition, the software should be secured if any failure happens. In the case of a failure state, it should fail in a way that it still protects it. Though the software is no longer functional, confidentiality and integrity can still be maintained. The software should also be designed to be user-friendly, even with security features implemented in a manner that it doesn't obstruct the GUI as users are likely to turn these off if authentication features in the app are invasive.

In the development stage, secure programming best practises provided by OWASP, can safeguard applications toward increased exploits. As a consequence, there would be no need to patch those bugs subsequently in the development cycle of the software, which lowers the maintenance and recovery costs. While doing code review, an extra layer of defence can be added to increase the security. Until it reaches the production stage, the code review stage should guarantee the software stability.

While performing penetration testing, the testing stage is usually based on identifying mistakes according to the specifications of the consumer. It is extremely important to verify the built software can withstand future cyberattacks. In relation, experimental penetration testing should also be done as the software reaches the release process in every instance of the SDLC with the help of developers before penetration testers start.

A collection of actions is given by Microsoft to assist developers after the end product is released [40]. First, every organization should develop an incident response plan by finding specific security contact information, or create a third-party code security support plan. Furthermore, the company should execute a definitive security examination to reveal vulnerabilities, while the final examination should ensure that all cases of misuse and security identified risks at the point of the requirements specification have been resolved. In addition, the organization should certify the end product and archive it. Credentialing it aims to ensure that all the programme specifications are fulfilled while archiving allows to execute further maintenance activities. Finally, organizations should always be ready to enforce the incidence response strategy if the software encountered any attacks.

## VI. Conclusion

In this paper, security protocols, procedures and methods that are currently being implemented for improving the security aspect of software have been reviewed. The information obtained from the field demonstrates the lack of consistent policies and guidelines in action at the level of project development at each point of the development process. Suggestions and verifications have been collected in this regard for classifying the actual practices that are reasonable and acceptable at each level of the software development life cycle. Interestingly, the findings show that the developers are not the only loopholes when it comes to the secure development process. In contrast to typical assumption, developers do not specifically neglect or disregard security, nor regard it as something external to their duty. They identify and associate with its significance, and they are motivated towards software security in many cases. On the other hand, the most critical software security countermeasures are mostly linked to the poor management of the procedure. Dealing with conflicting goals, the lack of security measures, policies, expertise, or infrastructure are the primary problems of security errors.

The purpose of this project is to examine the different approaches of secure software development to encompass information about cautious software development and make it available to developers who are not skilled in security by illustrating elements of the system architecture that could lead to exploits. In the preceding sections, we demonstrated how it is possible to use a secure development method to convey common computer security theories. Such descriptions analyze structures of different types and display fatal flaws. It is very crucial to address security from the very start of every software development process. However, it is not a simple process to establish the right security specifications for a project, and in many instances, due to their "non-functional" existence, requirements of this kind are difficult to track and evaluate throughout the development process. There are many efforts to encourage the task of identifying safety standards, but we have minimal understanding of the efficacy, and there is a deficiency of a thorough explanation of how to use them in practice. The vast majority were also built with conventional security-critical initiatives in consideration, which implies that, particularly for developers without adequate security knowledge, these techniques are not always suitable.

By proposing a solution that enables researchers to continue their development and validation in various environments, this study makes a significant contribution to awareness and practice. In addition, software engineers will be using the solution early in the design process to produce software that is more stable and well organized. As we know, numerous cases of software bugs and breaches exist towards the end of implementation cycles and these events are too expensive for the organization and the stakeholders to bear. For future work, it is crucial to discuss possible relationships between causes, deterrents, and software security techniques. Moreover, the findings from the discussion suggested that a security problem such as a security breach would increase the awareness of software security, which would be interesting to examine security mechanisms and behaviors in businesses that have encountered certain problems and compare it with those that haven't.